\documentclass[useAMS,usenatbib]{mn2e}
\usepackage{lscape}
\usepackage{amsmath,amssymb}
\usepackage{graphicx,mathptm} 
\usepackage{times} 
\usepackage{natbib}
\usepackage{psfig}
\usepackage{aas_macros}
\usepackage{xcolor}

\def\apjl{Astrophys. J. Lett.}
\def\apjs{Astrophys. J. Suppl.}

\def\aap{Astron. Astrophys.}

\newcommand{\Mpch}{$h^{-1}\,\mbox{Mpc}$\,}
\newcommand{\Gpch}{$h^{-1}\,\mbox{Gpc}$\,}

\title[3PCF in cDE models] {Disentangling interacting dark energy
cosmologies with the three-point correlation function}

\author[Moresco, Marulli, Baldi, Moscardini \& Cimatti] {Michele
  Moresco$^{1}$\thanks{E-mail: michele.moresco@unibo.it}, Federico
  Marulli$^{1,2,3}$, Marco Baldi$^{1,2,3}$, Lauro Moscardini$^{1,2,3}$
  \newauthor{and Andrea Cimatti$^{1}$} \\ $^1$Dipartimento di Fisica e
  Astronomia, Universit\`a di Bologna, Viale Berti Pichat 6/2, I-40127
  Bologna, Italy\\ $^2$INAF - Osservatorio Astronomico di Bologna, Via
  Ranzani 1, I-40127 Bologna, Italy\\ $^3$INFN - Sezione di Bologna,
  Viale Berti Pichat 6/2, I-40127 Bologna, Italy
}

\begin{document}

\maketitle

\begin{abstract}
We investigate the possibility of constraining coupled dark energy
(cDE) cosmologies using the three-point correlation function
(3PCF). Making use of the {\small CoDECS} N-body simulations, we study
the statistical properties of cold dark matter (CDM) haloes for a
variety of models, including a fiducial $\Lambda$CDM scenario and five
models in which dark energy (DE) and CDM mutually interact. We measure
both the halo 3PCF, $\zeta(\theta)$, and the reduced 3PCF,
$Q(\theta)$, at different scales ($2<r\,[$\Mpch$]<40$) and redshifts
($0\leq z\leq2$). In all cDE models considered in this work,
$Q(\theta)$ appears flat at small scales (for all redshifts) and at
low redshifts (for all scales), while it builds up the characteristic
V-shape anisotropy at increasing redshifts and scales. With respect to
the $\Lambda $CDM predictions, cDE models show lower (higher) values
of the halo 3PCF for perpendicular (elongated) configurations. The
effect is also scale-dependent, with differences between $\Lambda$CDM
and cDE models that increase at large scales. We made use of these
measurements to estimate the halo bias, that results in fair agreement
with the one computed from the two-point correlation function
(2PCF). The main advantage of using both the 2PCF and 3PCF is to
break the bias$-\sigma_{8}$ degeneracy. Moreover, we find that our
bias estimates are approximately independent of the assumed strength
of DE coupling. This study demonstrates the power of a
higher-order clustering analysis in discriminating between alternative
cosmological scenarios, for both present and forthcoming galaxy
surveys, such as e.g. BOSS and Euclid.
\end{abstract}

\begin{keywords} 
  cosmology: observations -- cosmology: theory -- dark energy -- dark
  matter -- large-scale structure of the Universe.
\end{keywords}


\section {Introduction}
\label{sec:intro}
The two-point correlation function (2PCF) of galaxies and galaxy
clusters, and its Fourier transform, the power spectrum, have been
proven to be extremely powerful tools to constrain cosmological
parameters. On one side, the shape and normalisation of the 2PCF, as
well as its redshift-space distortions due to galaxy peculiar
velocities, have been extensively exploited to estimate both
astrophysical and cosmological parameters, such as the mass density,
the growth rate of cosmic structures and the galaxy bias
\citep[e.g.][]{hawkins2003, daangela2005a, ross2007, guzzo2008,
zhang2008, cabre2009b, blake2011a, chuang2012, samushia2012,
sanchez2012, marulli2013, delatorre2013b}. On the other side, they
encode information on the primordial matter fluctuations in the form
of baryon acoustic oscillations (BAO), that have rapidly become one of
the standard cosmological probes to constrain dark energy (DE)
\citep[e.g.][]{eisenstein2005, cole2005, percival2007, percival2010,
sanchez2009, kazin2010, beutler2011, blake2011b, padmanabhan2012,
anderson2012, veropalumbo2013}.

The outcome of the above analysis highlighted the strong constraining
power of two-point statistic measurements that, together with
observations of Type Ia supernovae, weak lensing, and clusters of
galaxies, can be crucial to discriminate between alternative
cosmological scenarios \citep[e.g.][]{weinberg2013,
amendola2013}. This has driven the development of increasingly
larger photometric and spectroscopic redshift surveys, aimed at
reducing as much as possible the statistical uncertainties present in
these techniques. Many of these surveys are presently ongoing,
e.g. the Baryon Oscillation Spectroscopic Survey \citep[BOSS,
][]{boss}, the WiggleZ Dark Energy Survey \citep{blake2011a} and the
VIMOS Public Extragalactic Redshift Survey \citep[VIPERS,
][]{guzzo2013, garilli2013}, or are planned for the next future, such
as Euclid\footnote{http://www.euclid-ec.org/} \citep{euclid,
 amendola2013}.

In order to fully exploit the information encoded in the large scale
structure of the Universe, some theoretical and observational studies
have already started to go beyond the 2PCF and power spectrum,
considering also higher-order statistics, both in Fourier space,
i.e. using the bispectrum \citep[e.g.][]{fry1982, matarrese1997,
verde1998, scoccimarro2000, verde2000, scoccimarro2001,
sefusatti2005, sefusatti2007}, and in real space, i.e. using the
three-point correlation function (3PCF) \citep[e.g.][]{fry1994,
frieman1994, jing1997, jing2004, kayo2004, gaztanaga2005,
nichol2006, ross2006, kulkarni2007, marin2008, marin2011,
mcbride2011, mcbride2011b, marin2013, guo2014}. All these
previous analyses demonstrated the power of using higher-order
correlation measurements to explore the statistics of matter
distribution beyond the linear approximation, providing
supplementary information that may help in reducing degeneracies
between parameters. In particular, the 3PCF allows to investigate
how the spatial distribution of cosmic structures depends on
two-dimensional displacements, besides reciprocal distances, hence
representing the first significant statistical order to detect
non-Gaussian signals.
 
One of the main goals of the analyses cited above was to constrain the
linear galaxy bias, $b$. The advantage of estimating $b$ from the 3PCF
is that such a measurement is independent of the value of the power
spectrum normalisation $\sigma_{8}$, as we will extensively discuss in
the next sections. However, a standard $\Lambda$CDM cosmology is
generally assumed to model the 3PCF of the CDM component, and the
impact of this assumption on the estimated value of $b$ has never been
tested against alternative cosmological scenarios.

In this paper we measure the 3PCF of CDM structures in coupled dark
energy (cDE) models, using the largest N-body simulations to date of
these scenarios, i.e. the COupled Dark Energy Cosmological
Simulations \citep[{\small CoDECS}, see][]{CoDECS}. The spatial properties of
CDM haloes in the {\small CoDECS} simulations have already been
presented in \citet*{marulli2012a} (hereafter MBM12) and
\citet{cervantes2012}, using the 2PCF. In particular, these works
focused on the halo bias, on baryon acoustic oscillations (BAO), and
on the effects of the DE coupling on both geometric and dynamic
redshift-space distortions. The halo 2PCF in cDE models, and in
particular its redshift evolution, appears significantly different
with respect to the $\Lambda$CDM one. However, MBM12 also found a
strong degeneracy with $\sigma_{8}$, i.e. the effects of the DE
coupling can be mimicked by a $\Lambda$CDM cosmology with a rescaled
value of $\sigma_{8}$, or alternatively with a higher value of the total
neutrino mass \citep[e.g.][]{marulli2011, baldi2013}.

Extending the MBM12 analysis, we consider higher-order statistics of
CDM haloes, with the aim of breaking the $\sigma_{8}$-bias degeneracy. In
particular, we derive the halo bias from the reduced 3PCF, as a
function of redshift and scale and in different cDE models, and test
if a wrong assumption of the underlying cosmology can bias the
results.

The paper is organised as follows. In \S \ref{sec:cDE} we present the
{\small CoDECS} simulations, summarising their general properties. The
method and the algorithms implemented to estimate the 3PCF are
described in \S \ref{sec:3PCF}. In \S \ref{sec:results} we present our
results, discussing in particular the evolution of the 3PCF as a
function of redshift, scale, and cosmological model (\S
\ref{sec:growth}), comparing the halo bias factors estimated from the
2PCF and 3PCF (\S \ref{sec:bias}), and testing the impact of assuming
a wrong underlying cosmological model (\S
\ref{sec:real-analysis}). Finally, in \S \ref{sec:concl} we draw our
conclusions.


\section {Models and Simulations }
\label{sec:cDE}
The cDE cosmological scenario has been proposed as a viable alternative to the
standard $\Lambda$CDM cosmology, mainly to alleviate the
fine-tuning problems of the cosmological constant \citep{wetterich1995, amendola2000}. 
These models introduce an interaction between the DE scalar field, $\phi$, and the
CDM fluid. Such a coupling can be parameterised in different ways,
depending on the coupling function and the shape of the scalar self-interaction potential. In
this work we considered the cDE parameterisation proposed by
\citet{baldi2011a} and \citet{baldi2012}, where the coupling function $\eta (\phi )$ is 
parameterised as:
\begin{equation}
\eta (\phi)=\eta_{0}e^{\eta_{1}\phi} \,,
\end{equation}
where $\eta _{0}$ and $\eta _{1}$ are constant free parameters.
Then, following \citet{CoDECS}, we consider two different choices for the 
self-interaction potentials: an exponential potential
\citep{lucchin1985, wetterich1988}:
\begin{equation}
V(\phi)=Ae^{-\alpha\phi} \, ,
\end{equation}
and a SUGRA potential \citep{brax1999}:
\begin{equation}
V(\phi)=A\phi^{-\alpha}e^{\phi^{2}/2} \,,
\end{equation}
where $A$ is a non-negative constant and $\alpha $ is potential slope
parameter. A detailed analysis of the background evolution and of the
structure formation properties at linear and non-linear level in these
cosmologies can be found in \citet{baldi2011,baldi2011a,baldi2012}.
This class of cDE cosmologies has attracted significant interest
in the last decade as it allows to alleviate the fine-tuning of the
DE density, thanks to the presence of an early scaling solution
\citep[called the $\phi $-MDE solution, see][]{amendola2000}, where
the DE shares a constant fraction of the total energy budget of the
universe. Furthermore, the selective interaction between the DE
field and CDM particles, as originally suggested by
\citet{Damour_Gibbons_Gundlach_1990}, provides a way to avoid the
tight constraints on the strength of the associated fifth-force that
generically characterises modified gravity theories. In this
respect, cDE models represent one of the few classes of effectively
non-standard gravity that appear still consistent with present
observational data \citep[][]{Pettorino_2013} and should be targeted
by the next generation of wide astronomical surveys.

\begin{table}
\begin{center}
\caption{The cosmological models of the {\small CoDECS} suite, with their
main parameters.}
\begin{tabular}{llcccc}
\hline \hline
Model & Potential & $\alpha$ & $\eta_{0}$ & $\eta_{1}$ & $\sigma_{8}$\\
\hline
$\Lambda$CDM & $V(\phi)=A$ & -- & -- & -- & 0.809\\
EXP001 & $V(\phi)=Ae^{-\alpha\phi}$ & 0.08 & 0.05 & 0 & 0.825\\
EXP002 & $V(\phi)=Ae^{-\alpha\phi}$ & 0.08 &  0.1 & 0 & 0.875\\
EXP003 & $V(\phi)=Ae^{-\alpha\phi}$ & 0.08 &  0.15 & 0 & 0.967\\
EXP008e3 & $V(\phi)=Ae^{-\alpha\phi}$ & 0.08 &  0.4 & 3 & 0.895\\
SUGRA003 & $V(\phi)=A\phi^{-\alpha}e^{\phi^{2}/2}$ & 2.15 & -0.15 & 0 & 0.806\\
\hline \hline
\end{tabular}
\label{tab:cDE}
\end{center}
\end{table}

For the purpose of our analysis, we make use of the {\small CoDECS}
simulations \citep{CoDECS}. In particular, we analyse the publicly available
{\small L-CoDECS} sets\footnote{Public catalogues available at 
http://www.marcobaldi.it/CoDECS}, that are collisionless N-body simulations of
$2\times 1024^{3}$ particles for the (coupled) CDM and (uncoupled) baryon 
fields, in a periodic cosmological box of $1$ \Gpch per side. The {\small CoDECS} 
simulations consider six different cosmological models, consisting of five cDE cosmologies and
one standard $\Lambda$CDM model, assumed as the fiducial reference. The main parameters
of these models are reported in Table \ref{tab:cDE}. All the {\small CoDECS} simulations have 
been generated with initial conditions obtained from the same
linear power spectrum at $z_{\rm CMB}\approx 1100$.
At $z=0$, all the models assume the following
background cosmological parameters: $H_{0}=70.3\; {\rm km s^{-1} Mpc^{-1}}$,
$\Omega_{\rm CDM}=0.226$, $\Omega_{\rm DE}=0.729$,
$\Omega_{b}=0.0451$, consistent with WMAP7 data
\citep{komatsu2011}. However, as a consequence of the new physics associated 
to the coupling, the different models are characterised by different values of $\sigma _{8}$ 
at the present time (see again Table~\ref{tab:cDE}).

In order to build the {\small CoDECS} public halo catalogues that are
used in this analysis, CDM haloes have been identified using a
Friend-of-Friend algorithm \citep[FoF, ][]{davis1985}, with linking
length $\lambda=0.2\,\bar{d}$, where $\bar{d}$ is the mean CDM
interparticle separation. Baryon particles have then been attached to
the FoF group of their nearest CDM neighbour. Finally, gravitationally
bound substructures have been identified with the {\small SUBFIND}
algorithm \citep{springel2001}. To be consistent with MBM12, we use
mass-selected sub-halo catalogues, with masses in the ranges $M_{\rm
min} < M < M_{\rm max}$, where $M_{\rm min} =
2.5\cdot10^{12}\,h^{-1}\,M_{\odot}$ and $M_{\rm max}=3.6\cdot10^{15}$,
$1.1\cdot10^{15}$, $4.9\cdot10^{14}$, $2.6\cdot10^{14}$,
$1.8\cdot10^{14}\,h^{-1}\,M_{\odot}$ at $z=0, 0.55, 1, 1.6, 2$,
respectively. By applying the same mass cut as in MBM12, we are
able to properly quantify the impact of higher-order statistics with
respect to lower-order ones. Nevertheless, this assumption does not
impact the results presented in this paper.


\section {The three-point correlation function}
\label{sec:3PCF} 
The probability of finding triplets of objects at relative comoving
distances $r_{12}$, $r_{23}$, and $r_{31}$ can be written as:
\begin{multline}
d{\rm
  P}=\bar{n}^{3}[1+\xi(r_{12})+\xi(r_{23})+\xi(r_{31})+\zeta(r_{12},r_{23},r_{31})]
\\ dV_{1}dV_{2}dV_{3} \, ,
\end{multline}
were $\bar{n}$ is the average density of objects, $V_i$ are comoving
volumes, and $\xi$ and $\zeta$ are the 2PCF and the 3PCF, respectively
\citep{peebles1980}.

The choice of the shape of triangles is not unique. A standard
method is to consider equilateral triangles, so that the 3PCF will
depend on one variable only, that is the scale (i.e. the triangle side). In this
analysis we adopt a different strategy, which is also widely used in literature: 
we fix two sides of the triangles and vary the angle, $\theta$, between them. In this
configuration, the angles $\theta\sim0$ and $\theta\sim\pi$ represent
the {\em elongated} configurations, while $\theta\sim\pi/2$ is the
{\em perpendicular} configuration \citep{gaztanaga2005}. This choice
allows us to study, at the same time, both the scale (by varying the length of the first two 
sides) and the shape (by varying the angle between the first two sides)
dependence of the 3PCF, maximizing the amount of information that can
be extracted. To parameterise the triangles formed by triplets of
objects, we adopt the definition given by \citet{marin2011}:
\begin{equation}
  \left\{
    \begin{array}{l}
      r_{12}\\
      r_{13}=u\; r_{12}\\
      r_{23}=r_{12}\;\sqrt{1+u^{2}-2\; u \; \cos\theta} \nonumber 
    \end{array}
    \right.
\end{equation}
with a constant logarithmic binning in $\Delta r_{ij}/r_{ij}$. This
binning scheme is useful to include triangles with similar shapes in
each $\theta$-bin (but see e.g. \citet{nichol2006} and
\citet{kulkarni2007}, for a different approach), and has been demonstrated
to better reproduce theoretical prediction, and to have smaller associated
errors compared to other paramererisations \citep[for a detailed discussion, we refer to][]{marin2011}.

While the values of $\zeta(\theta)$ at different angles can vary by
orders of magnitude, depending on the scale considered, the reduced
3PCF $Q$, defined as:
\begin{equation}
Q(r_{12},r_{13},\theta) \equiv
\frac{\zeta(r_{12},r_{23},\theta)}{\xi(r_{12})\xi(r_{23})+\xi(r_{23})\xi(r_{31})+\xi(r_{31})\xi(r_{13})}
\, ,
\end{equation}
exhibits less variations with scale, as it can be shown that
$\zeta\propto\xi^{2}$ in hierarchical scenarios \citep{peebles1975}.

In this paper, we measure the 3PCF using the
\citet{szapudi1998} estimator. For a data sample of $N_{D}$ elements and a corresponding
random catalogue of $N_{R}$ elements, such estimator allows to compute the 3PCF as:
\begin{equation}
\zeta(r_{12},r_{12},\theta) = \frac{DDD-3DDR+3DRR-RRR}{RRR} \, ,
\end{equation}
where $DDD$, $RRR$, $DDR$, and $DRR$ are the numbers of data triplets,
random triplets, data-data-random triplets, and data-random-random
triplets, normalised by $N_{D}^{3}/6$, $N_{R}^{3}/6$,
$N_{D}^{2}N_{R}/2$, and $N_{D}N_{R}^{2}/2$, respectively.

The 2PCF is calculated using the standard \citet{landy1993} estimator:
\begin{equation}
\xi(s) = \frac{DD-2DR+RR}{RR} \, ,
\end{equation}
where $DD$, $RR$ and $DR$ are the numbers of data pairs, random pairs,
and data-random pairs, normalised by $N_{D}^{2}/2$, $N_{R}^{2}/2$, and
$N_{D}N_{R}$, respectively.

\begin{figure*}
\includegraphics[angle=-90, width=1.0\textwidth]{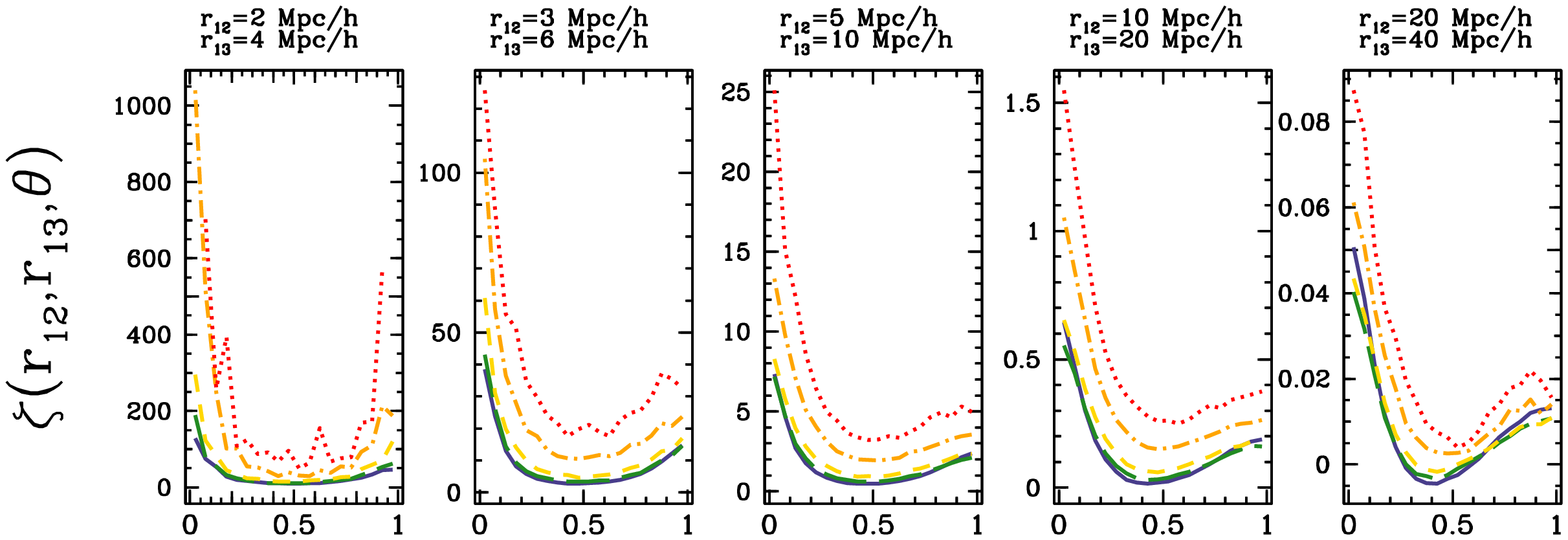}
\includegraphics[angle=-90, width=1.0\textwidth]{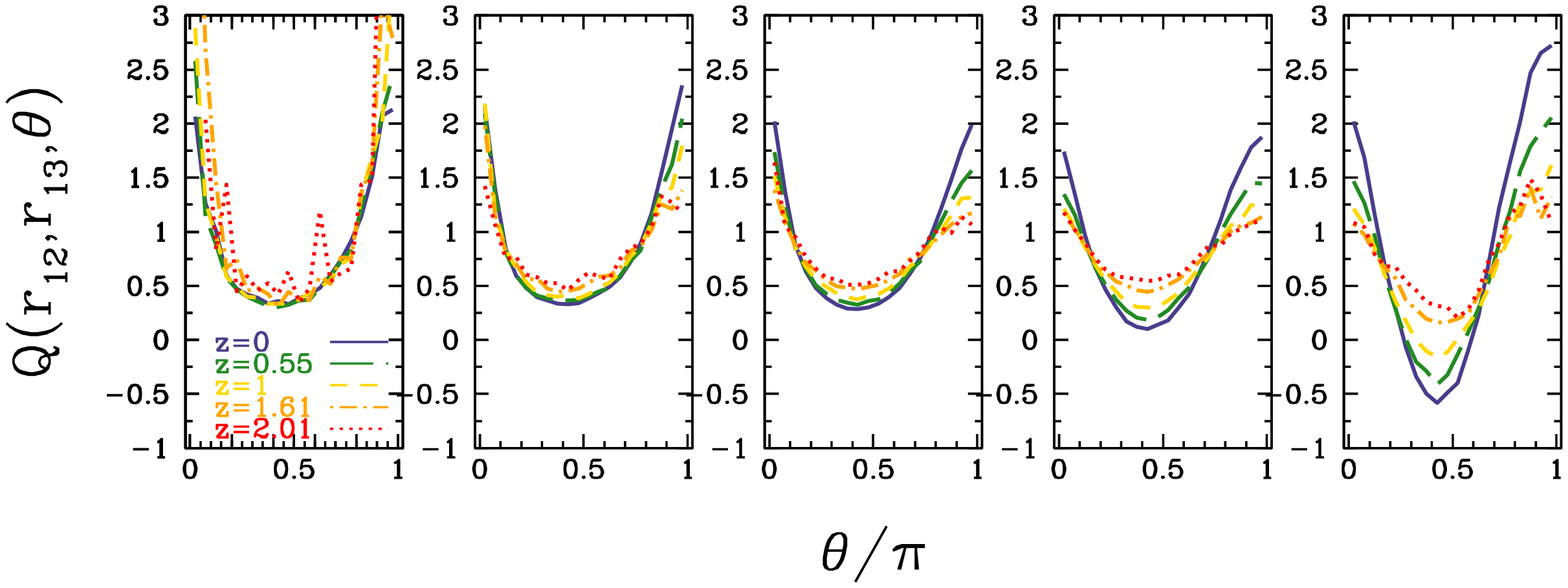}
\caption{The halo 3PCF $\zeta(\theta)$ (upper panels) and reduced 3PCF
$Q(\theta)$ (lower panels) in the $\Lambda$CDM cosmology, as a
function of redshift (coloured curves) and for different scales
(panels from left to right), as indicated by the labels in the upper
part of each panel.}
\label{fig:LCDM}
\end{figure*}

To measure the 3PCF, we implemented a {\it linked-list} based
algorithm, extending a preexisting numerical code for the computation
of the 2PCF \citep{marulli2012a, marulli2012b, marulli2013}. The
sketch of the algorithm is the following: (i) both the data and random
catalogues are divided into sub-regions; (ii) the indexes of the
objects in each sub-region are stored in a {\em linked} vector; (iii)
for each object, only the sub-regions close to it (up to a maximum
scale) are considered; iv) the objects in these close sub-regions are
retrieved with a fast search inside the {\em linked} vector. In this
way, we avoid taking into account the regions that would not contribute
to the triplets at the desired scale. Thanks to the {\it linked-list}
approach, the code results to be extremely fast in measuring both the
2PCF and the 3PCF. As we have directly verified, the way the
sub-regions are defined in the {\it linked-list} method does not
impact the code outputs, but only its performances. This technique
allows us to save a significant amount of computational time,
depending on the scale considered. For instance, with $1.5\cdot10^6$
objects in a box of $1$ \Gpch per side, and considering the case $u=2$
\Mpch, the {\it linked-list} method can reduce the number of
operations by a factor of $\sim8\cdot10^{8}$ for $s=10$ \Mpch, and up to
$\sim10^{13}$ for $s=2$ \Mpch.


\section{Results}
\label{sec:results}

\subsection{The growth of structure in cDE models}
\label{sec:growth}
All the cDE models analysed in this paper are characterised by a background
expansion history that does not deviate more than $\sim 6\%$ from the evolution 
of the reference $\Lambda$CDM reference cosmology. 
Therefore, they are nearly indistinguishable from $\Lambda $CDM in terms of the
Hubble function $H(z)$, especially at low redshifts and to discriminate between them it is
necessary to look at the growth of structures, both in the linear and in the
non-linear regimes.

To investigate the evolution of CDM haloes in the {\small CoDECS}
simulations, we measure both the 3PCF, $\zeta(\theta)$, and the
reduced 3PCF, $Q(\theta)$. We perform our analysis at five redshifts,
$z=[0, 0.55, 1, 1.61, 2.01]$, and at different scales, fixing $u=2$
\Mpch and $r_{12}=2,3,5,10,20$ \Mpch. In this way, we are able to probe the
properties of CDM structures from small to intermediate
scales. We adopt an angular binning of $\Delta\theta=\pi/20$, to
follow in detail the shape dependence of both $\zeta(\theta)$ and
$Q(\theta)$, and $\Delta r_{ij}/r_{ij}=0.1$ as a trade-off between
larger values, that would increase the covariance between different
bins, and smaller values, that would decrease the signal-to-noise
ratio of our measurements. We perform two types of analysis. On one
side, we measure the 3PCF for each cosmological model, to study its
redshift and space dependence. On the other side, we compare each
cosmological model with the $\Lambda$CDM one, looking for possible
differences.

\begin{figure*}
\includegraphics[angle=-90, width=0.95\textwidth]{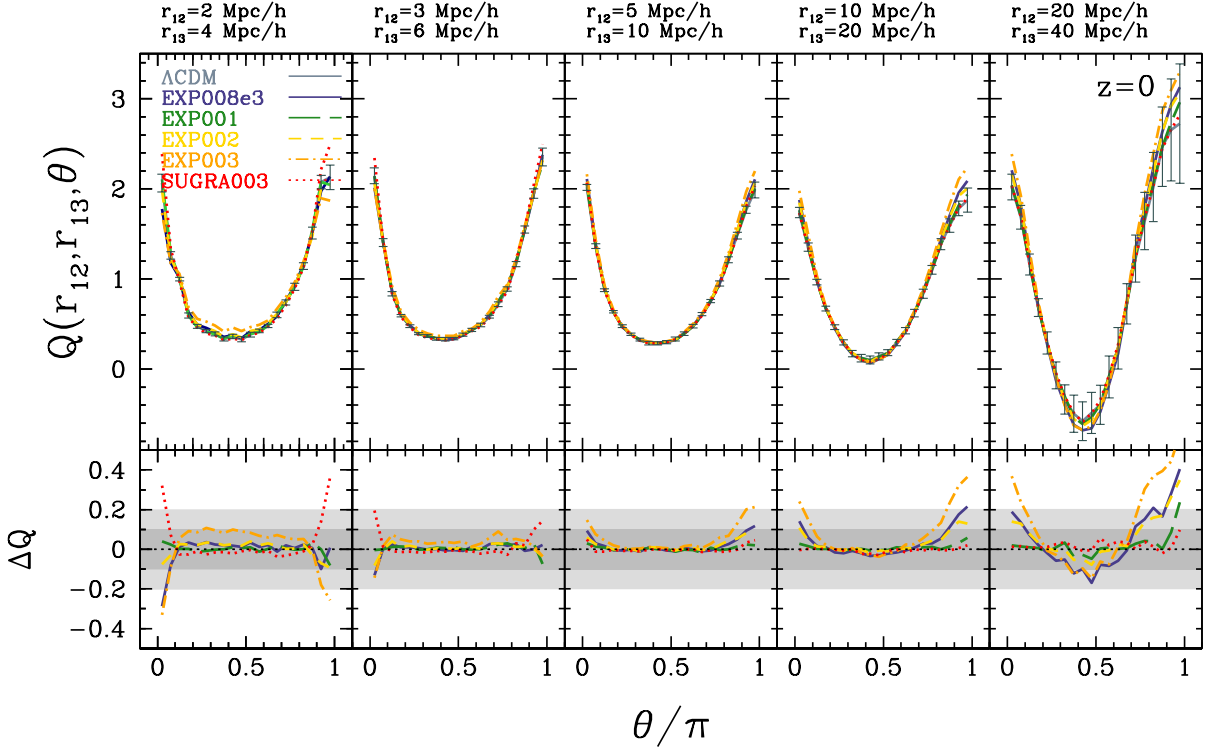}
\caption{The reduced 3PCF $Q(\theta)$ of CDM haloes at $z=0$, for
$\Lambda$CDM and cDE cosmologies (coloured curves) and for different
scales (panels from left to right), as labeled in the upper part of
each panel. The errorbars (shown in dark grey) have been obtained as
the scatter among the 3PCF measured in sub-volumes of the larger
{\small XL-CoDECS} simulation. Light and dark shaded areas in the
lower plots show differences $\Delta Q<$0.1and 0.2, respectively}
\label{fig:z0}
\end{figure*}

\begin{figure*}
\includegraphics[angle=-90, width=0.95\textwidth]{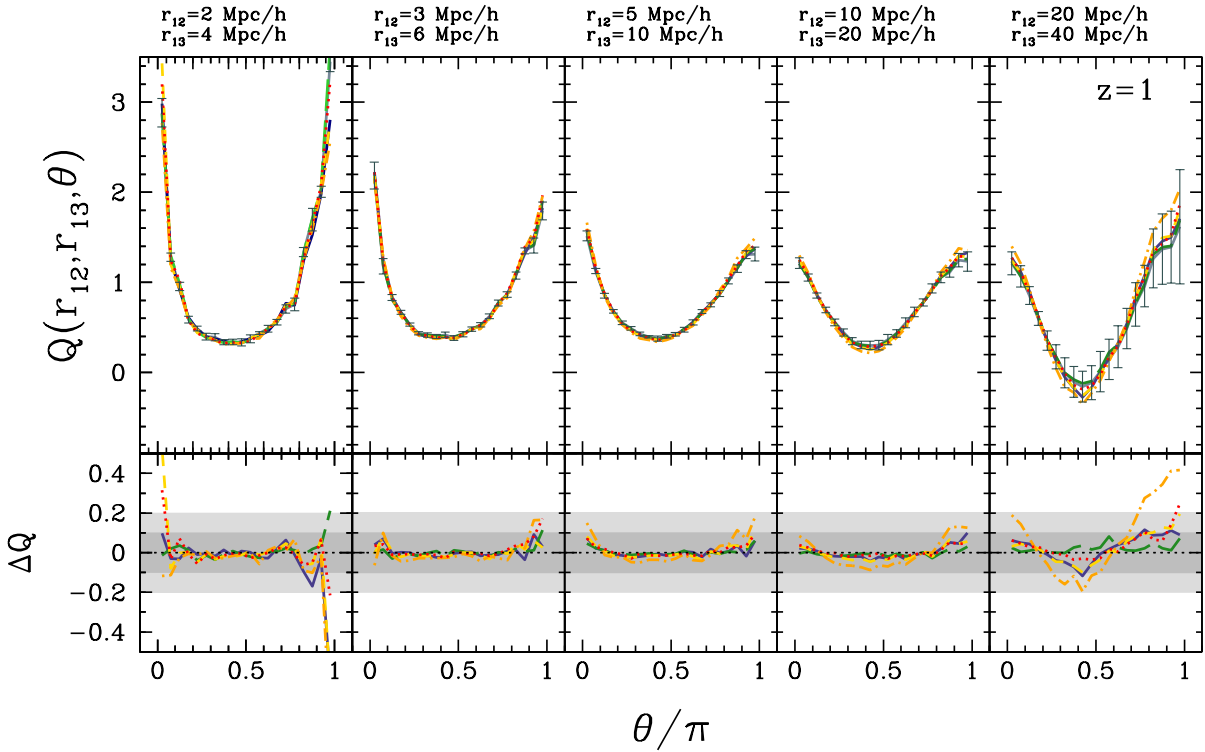}
\caption{Same as Fig. \ref{fig:z0}, but for $z=1.$}
 \label{fig:z1}
\end{figure*}

\begin{figure*}
\includegraphics[angle=-90, width=1.0\textwidth]{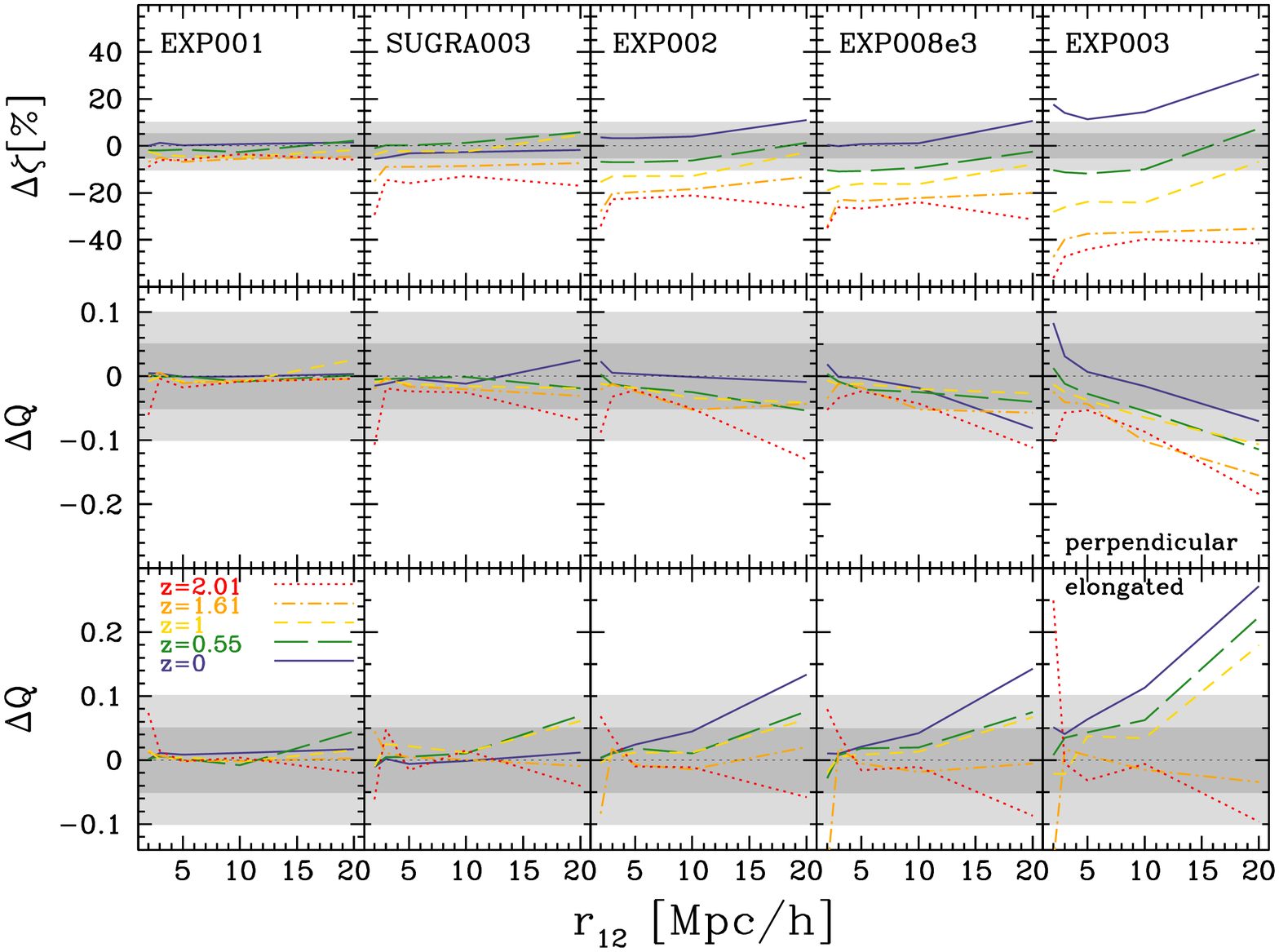}
\caption{Mean fractional differences in percent $\Delta\zeta[\%]$
(upper panels) and mean differences $\Delta Q$ (lower panels) of
cDE models with respect to the $\Lambda$CDM case (panels from left
to right), as a function of redshift (different colours). The
variation $\Delta Q$ has been estimated for both perpendicular
(central panels) and elongated configurations (lower panels). Light
(dark) shaded areas show the 10\% (5\%) levels in the upper
panels, and $\Delta Q<$0.1 (0.05) in the central and lower panels.
The quantity reported on the x-axis is the length of the first side
of the triangle, $r_{12}$.}
 \label{fig:modelcomparison}
\end{figure*}

\subsection{Redshift and scale dependence}
Figure~\ref{fig:LCDM} shows the 3PCF $\zeta(\theta)$ (upper panels) and
$Q(\theta)$ (lower panels), as a function of redshift and scale, as
indicated by the labels. Here we show only the $\Lambda$CDM results,
since the trends we are about to discuss are similar in the different
models. The main results of this first analysis can be summarised as
follows:

\begin{enumerate}
\item at each scale and redshift, the reduced 3PCF $Q$ is higher for
elongated triangles than for perpendicular ones. This is a well
understood effect \citep{gaztanaga2005, marin2008}, due to the
fact that cosmic structures preferentially move along gradients of
the density field within non-linear gravitational instabilities
\citep{bernardeau2002};
\item at fixed scale, we find a clear redshift trend in both $\zeta$
and $Q$: the halo 3PCF increases going to high redshift, while the
reduced 3PCF becomes flatter. The former result simply reflects the
evolution of the bias function in mass-selected samples (see also \S
\ref{sec:bias}), while the evolution of the reduced 3PCF can be
interpreted as the imprint of the formation of filaments along the
cosmic time. Indeed, this trend is more evident at large scales, 
with differences up to $\Delta Q\sim 1$ between $z\sim0$ and $z\sim2$
We also notice that the difference of $Q$ between elongated and perpendicular
configurations increases with decreasing redshift;
\item at fixed redshift, the reduced 3PCF exhibits a transition from a
{\it U-shape}, at small scales, to a {\it V-shape}, at large scales,
consistently with previous results from numerical investigations
\citep[see][]{gaztanaga2005, marin2008}. This is a consequence of
the fact that, at small scales, structures reside preferentially in
rounder structures (hence $Q$ is flatter), while at larger scales
the contribution of structures in filaments starts becoming
important.
\end {enumerate}

\subsection{Impact of different cosmological models}
As a next step, we compare the 3PCF of different cosmological models,
at fixed scale and redshift. The result is shown in Figs.~\ref{fig:z0} and \ref{fig:z1}
at redshifts $z=0$ and $z=1$ (respectively),
as examples. In the lower panel of each plot we show the
difference, $\Delta Q$, between the values of $Q$ of each cosmological
model and the $\Lambda$CDM ones. The errorbars are estimated from the
scatter among the 3PCF measured in sub-volumes of a larger
simulation. Specifically, we used the CDM sub-halo catalogues extracted
from the new {\small XL-CoDECS} simulations, generated with the same
pipeline described in \S \ref{sec:cDE}, but in a larger volume of $(2
h^{-1}\,\mbox{Gpc})^3$. We divided each snapshot of the $\Lambda$CDM
{\small XL-CoDECS} simulation into $27$ sub-cubes, and used them to
estimate the full normalised covariance matrix of our
measurements. Each sub-cube is $\sim 700$ \Mpch on a side, so its
volume is comparable to the one of the {\small L-CoDECS}
simulations. The errorbars shown in Figs.~\ref{fig:z0} and \ref{fig:z1} are the
r.m.s. between the 3PCF measured in these sub-cubes, i.e. the square
root of the diagonal elements of the covariance matrix. More details
can be found in Appendix \ref{sec:cov}.

\begin{figure*}
\includegraphics[angle=0, width=1.0\textwidth]{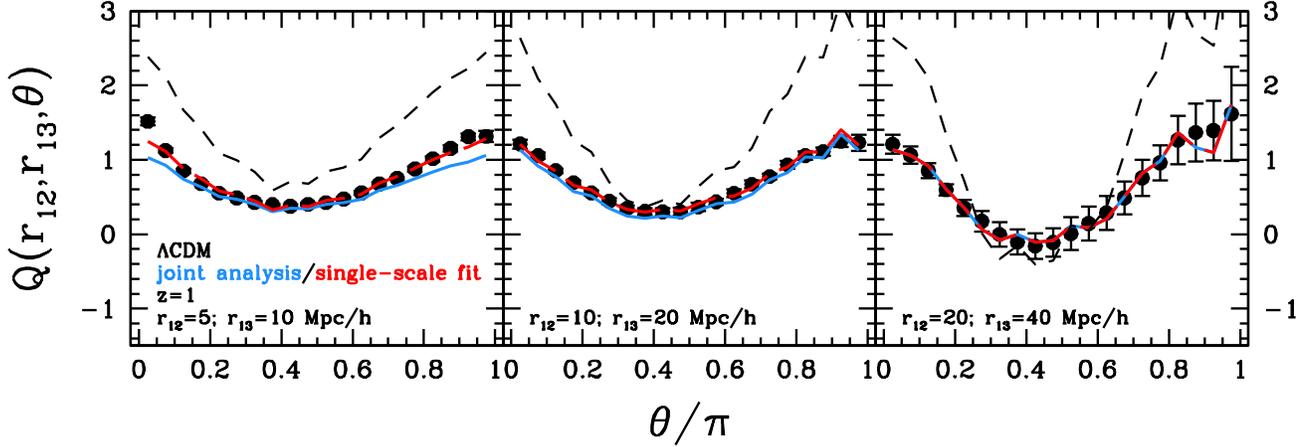}
\caption{The bias estimated from the 3PCF in the $\Lambda$CDM
simulation at $z=1$. The reduced 3PCF of DM {\bf(dashed
lines)} and haloes (points with errorbars) is shown at different
scales $\{r_{12},r_{13}\}=\{5,10\},\{10,20\},\{20,40\}$ \Mpch,
from left to right. The long-dashed red lines and the solid
blue lines show the best-fit values of $Q(\theta)$ obtained with
Eq.~\ref{eq:bias3PCF} from a fit to each single scale and from a
joint analysis, respectively.}
\label{fig:biasLCDM_z1}
\end{figure*}

In Fig.~\ref{fig:modelcomparison} we report the mean fractional
differences in percent in $\zeta$ (upper panels) and the mean
differences $\Delta Q$ for perpendicular (intermediate panels) and
elongated configurations (lower panels), between cDE and $\Lambda$CDM
predictions. The latter quantities are obtained by averaging the
differences at each scale over the angle $\theta$, and are reported as
a function of the length of the first side of the triangles,
$r_{1}$. We verified that the differences in $\zeta$ do not vary
significantly as a function of the angle, presenting a shift that is
almost constant as a function of scale. Therefore, we decided to
average those differences over the full range
$0\leq\theta\leq\pi$. On the other hand, being normalised over the
2PCF, the differences in $Q$ averaged over the entire angle range
present an almost null shift, while the angular dependence is more
significant. Hence, we considered the cases of perpendicular and
elongated configurations separately, averaging the values of $\Delta
Q$ in the ranges $0.3<\theta/\pi<0.7$ and $\theta/\pi\leq0.3\; \cup\;
\theta/\pi\geq0.7$, respectively. For illustrative purpose, in
Tab.~\ref{tab:diff} we reported the mean fractional differences for
the EXP003 model at $z=1$.

\begin{table}
\begin{center}
\caption{Mean fractional differences in percent $\Delta\zeta[\%]$ and
mean differences $\Delta Q$ for EXP003 model at $z=1$.}
\begin{tabular}{lccc}
\hline \hline
scale & $\Delta\zeta[\%]$ & $\Delta Q$ & $\Delta Q$\\
$\rm [h^{-1}Mpc]$  &  & perpendicular & elongated\\
\hline
2-4 & 7.5 & -0.01 & -0.02\\
3-6 & -28 & -0.02 & -0.02\\ 
5-10 & -26 & -0.04 & 0.04 \\
10-20 & -23.8 & -0.06 & 0.03 \\
20-40 & -6.8 & -0.11 & 0.18\\
\hline \hline
\end{tabular}
\label{tab:diff}
\end{center}
\end{table}

The most interesting findings shown in Figs.~\ref{fig:z0} and \ref{fig:z1} and
\ref{fig:modelcomparison} can be summarised as follows:
\begin{enumerate}
\item at fixed redshift, we find significant differences between the
halo 3PCF of cDE and $\Lambda$CDM simulations, that become more and more
evident at increasing scales;
\item cDE models display a higher (lower) reduced 3PCF, with respect
to the $\Lambda$CDM scenario, for elongated (perpendicular)
configurations;
\item at fixed scale, the differences are larger for elongated
configurations;
\item the differences increase with redshift, being almost
negligible at $z=0$. 
\end{enumerate}

While the differences between the $\zeta$ values in $\Lambda$CDM and
cDE simulations are independent of the considered scale, there is a
strong shape dependence in $\Delta Q$, which presents opposite trends
as a function of the configuration: cDE cosmologies have always a
smaller (larger) 3PCF than $\Lambda$CDM for perpendicular (elongated)
triangles. The differences in $\zeta$ between cDE and $\Lambda$CDM
simulations are in line with the 2PCF measurements by MBM12.

We conclude that both the normalisation and the shape of the 3PCF
of cDE models are different from the $\Lambda$CDM ones, and can be
used to discriminate among models. Only the EXP001 model appears
almost indistinguishable from $\Lambda$CDM in terms of the 3PCF, in
the whole range of redshifts and scales considered. The model that
deviates the most from the $\Lambda$CDM cosmology is EXP003, with
differences up to $\sim 50\%$ in $\zeta$, at $z\sim2$.


\subsection{From the 3PCF to the halo bias}
\label{sec:bias}

\begin{figure}
\includegraphics[angle=0, width=0.43\textwidth]{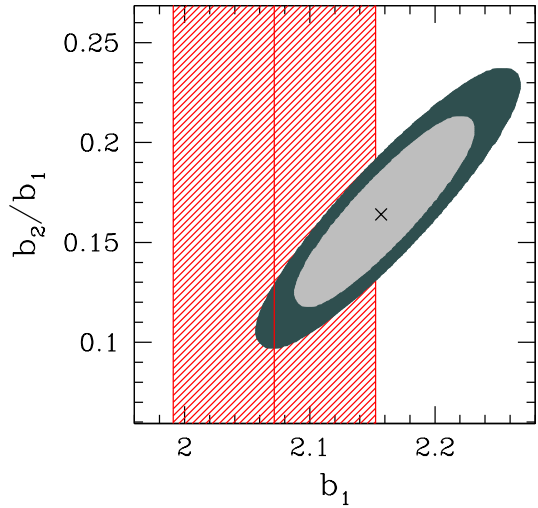}
\caption{Contour plot of $b_{1}$ and $b_{2}/b_{1}$ estimated at
$z=1$ from the joint analysis presented in
Fig. \ref{fig:biasLCDM_z1}. Light and dark shaded area represent
the 68\% and 95\% confidence levels. The vertical line indicates
the value of $b_{1}$ estimated with Eq.~\ref{eq:bias2PCF}, while
the red shaded area is the associated $1\,\sigma$ uncertainty.}
\label{fig:biasLCDM_z1_cont}
\end{figure}

From the measured 2PCF of a given sample of astrophysical sources it
is possible to infer their bias function with respect to the
underlying DM distribution. Using a Taylor expansion of the halo biasing 
function to the second order \citep[][]{fry1993}, the halo
overdensity, $\delta_{\rm h}$, can be expressed, at large scales, as a
function of the DM overdensity, $\delta_{\rm DM}$, as follows:
\begin{equation}
\delta_{\rm h}\approx b_{1}\delta_{\rm DM}+\frac{b_{2}}{2}\delta^{2}_{\rm DM} \, ,
\label{eq:deltah}
\end{equation}
where $b_{1}$ and $b_{2}$ are two bias factors. From
Eq.~\ref{eq:deltah}, we can derive a simple relation between the halo 2PCF
and the linear halo bias:
\begin{equation}
\xi_{\rm h}(r)\approx b_{1}^{2}\xi_{\rm DM}(r) \, ,
\label{eq:bias2PCF}
\end{equation}
where $\xi_{\rm DM}$ and $\xi_{\rm h}$ are the 2PCF of DM and haloes,
respectively. A similar relation can be derived also for the 3PCF:
\begin{equation}
Q_{\rm h}(\theta)\approx\frac{1}{b_{1}}\left(Q_{\rm DM}(\theta)+\frac{b_{2}}{b_{1}}\right) \, ,
\label{eq:bias3PCF}
\end{equation}
where $Q_{\rm DM}$ and $Q_{\rm h}$ 
are the reduced 3PCF of DM and haloes, respectively.

\begin{figure*}
\includegraphics[angle=0, width=1.0\textwidth]{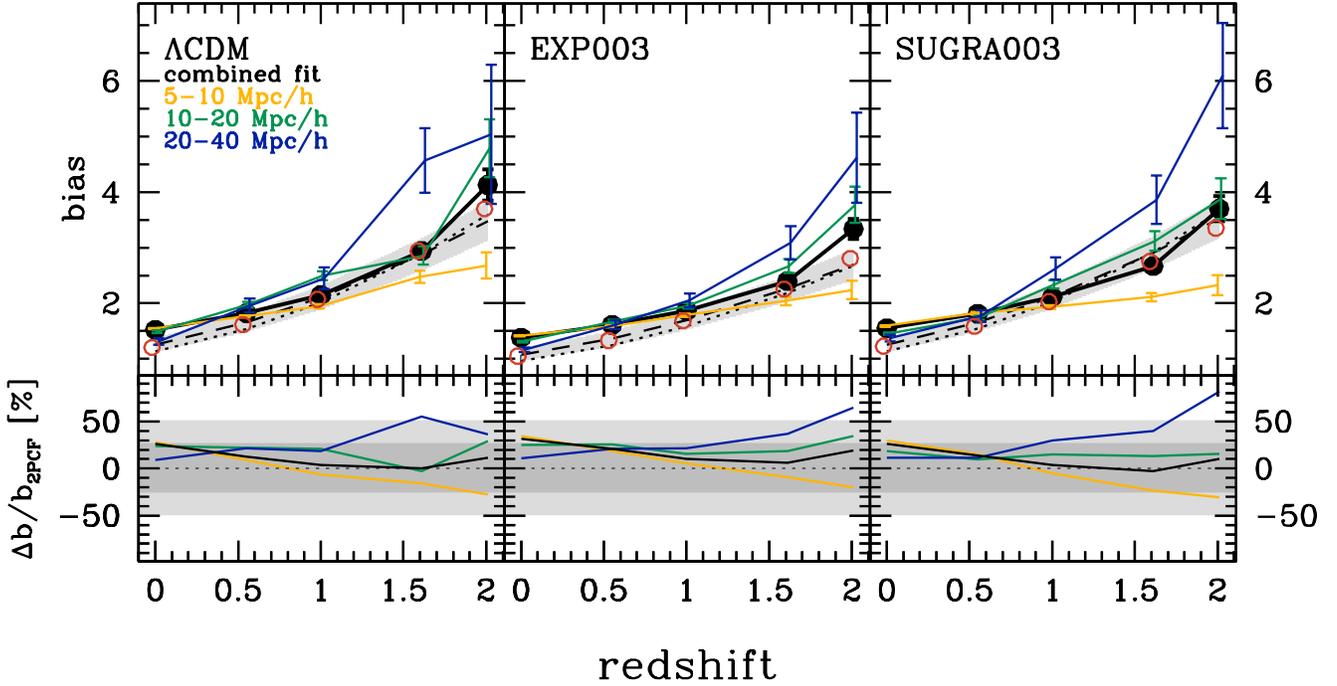}
\caption{Bias of CDM haloes estimated from the 3PCF
(Eq.~\ref{eq:bias3PCF}) in three different cosmological models.
{\em Upper panels:} the coloured lines show the best-fit values
obtained for $r_{12}=5, 10, 20$ \Mpch and $r_{13}=2r_{12}$, and by
combining together all the scales (yellow, green and blue lines, and
black dots, respectively). Open red dots show the bias as estimated
from the 2PCF (Eq.~\ref{eq:bias2PCF}) in the range
$10<r[$\Mpch$]<50$. The dashed and dotted lines display the
theoretical effective bias as predicted by
\citet{sheth_mo_tormen2001} and \citet{tinker2010}, with the grey
shaded areas showing a $10\%$ error, representative of theoretical
uncertainties. {\em Lower panels:} fractional difference in
percent between the bias estimated from 3PCF and 2PCF, defined as
$\Delta b/b_{2PCF}[\%]=(b_{3PCF}-b_{2PCF})/b_{2PCF}\cdot100$.
The coloured lines are the same as in the upper panels. The light
and dark shaded areas represent the 25\% and 50\% levels. The points have
been slightly shifted along the horizontal axis to make the figure clearer.}
\label{fig:bias_nocov2}
\end{figure*}

Different cosmological models predict different values for $b_{1}$ and
$b_{2}/b_{1}$, and a distinct redshift evolution. Therefore, the halo bias
can be a powerful probe to discriminate between alternative
cosmological scenarios. In particular, it can be shown that the bias
in cDE models is significantly different than the one predicted by the
$\Lambda$CDM scenario (see e.g. MBM12).

However, two key issues arise when analysing real datasets. Firstly,
the halo bias has to be inferred from the measured galaxy bias, that
depends not trivially on both baryon phenomena and selection
effects. Secondly, if the bias is estimated from the 2PCF, i.e. using
Eq.~\ref{eq:bias2PCF}, a fiducial value of $\sigma_{8}$ has to be
assumed to compute $\xi_{\rm DM}$. Since $\xi_{\rm
 DM}\propto\sigma_{8}^{2}$ at large scales, the bias amplitude scales
approximately as $\sigma_{8}^{-1}$ in the linear regime. MBM12 found
that the suppression of the halo bias caused by the DE coupling is
degenerate with $\sigma_{8}$ at scales $r\gtrsim 5$ \Mpch, i.e. the
different halo biases of cDE cosmologies may be simply recovered with
a $\Lambda$CDM model with a rescaled value of $\sigma_{8}$. Therefore,
without any prior on $\sigma_{8}$, it is impossible to detect any
signature of DE coupling using only the 2PCF at large scales.

The same considerations hold for the 3PCF as well, since it can be
shown that the matter 3PCF, $\zeta_{\rm DM}$, scales as
$\sigma_{8}^{4}$ \citep{pan2005}. On the contrary, the reduced 3PCF
$Q\sim\zeta/\xi^{2}$ does not depend on $\sigma_{8}$ by construction,
so that the bias factors estimated with Eq.~\ref{eq:bias3PCF} do not
require any prior on $\sigma_{8}$.

To derive the bias factors $b_{1}$ and $b_{2}/b_{1}$ of
Eq.~\ref{eq:bias3PCF}, we apply a standard $\chi^{2}$ minimization
approach using only the diagonal elements of the covariance matrix,
i.e. minimizing the function:
\begin{equation}
\chi^{2}=\sum \frac{(Q_{\rm h}-Q_{\rm h}^{\rm model})^{2}}{\sigma_{Q}^{2}} \, ,
\label{eq:chi}
\end{equation}
where the errors $\sigma_{Q}$ have been estimated as discussed in \S
\ref{sec:growth}, and $Q_{\rm h}$ and $Q_{\rm h}^{\rm model}$ are the
measured and theoretical reduced 3PCFs, respectively, the latter
quantity being given by Eq.~\ref{eq:bias3PCF}, where the CDM reduced 3PCF 
$Q_{\rm DM}$ is estimated using sparsely sampled sub-sets of
the CDM particle snapshots, to decrease the computational time. The
uncertainties estimated through Eq.~\ref{eq:chi} are slightly
underestimated, not taking into account the full covariance between
measurements in different bins, as discussed in Appendix
\ref{sec:cov}, where we also discuss the impact of considering the
full covariance matrix in the analysis. However, for the purpose of this work this is
not relevant: all the {\small CoDECS} simulations match the same
cosmology at $z_{\rm CMB}$ by construction, so the differences between
the reduced 3PCF measured in these catalogues are {\em real}
differences. The aim of showing uncertainties in the bias factors is
just to give a qualitative feeling on the expected uncertainties in
surveys comparable to the {\small CoDECS} samples in terms of volume
and density of objects.

Equations~\ref{eq:bias2PCF} and \ref{eq:bias3PCF} are reliable
approximations of the halo clustering only at large, linear
scales. Therefore, we use them to derive the halo bias at our three
largest configurations, in which triangles have sides
$\{r_{12},r_{13}\}=\{5,10\},\{10,20\},\{20,40\}$ \Mpch,
respectively. Different scales have been examined both separately and
jointly.

As an illustrative case, we show the result of the analysis for the
$\Lambda$CDM model at $z=1$. In Fig.~\ref{fig:biasLCDM_z1} is
presented the reduced 3PCF of CDM (dotted lines) and haloes (points
with errorbars) at the different scales. Red and blue lines display
the best-fit models for the reduced 3PCF of haloes, obtained from
Eqs.~\ref{eq:bias3PCF} considering each scale separately and with a
joint analysis, respectively. The best fit obtained by fixing the bias
with the 2PCF from Eq.~\ref{eq:bias2PCF} is not shown, since it
perfectly overlaps with the fit obtained with the 3PCF joint
analysis. In this case, the $b_{2}$ parameter has been obtained
 by minimizing the $\chi^{2}$, having fixed the bias $b_{1}$. The
complete agreement between the latter, at all scales, demonstrates
that the linear halo bias factors derived from the 2PCF and from the
3PCF are equivalent, at least at the redshifts and scales considered
here. In summary, Fig.~\ref{fig:biasLCDM_z1} shows that the method to
derive the halo bias from the 3PCF is reliable at $z=1$ and at the
scales considered here. As already noted before, the crucial advantage
of using the reduced 3PCF $Q(\theta)$ is that it does not depend on
$\sigma_{8}$, differently from the 2PCF and the 3PCF
$\zeta(\theta)$.\\ This can also be appreciated in
Fig.~\ref{fig:biasLCDM_z1_cont}. It shows the bias parameters $b_{1}$
and $b_{2}/b_{1}$, estimated from the 3PCF (grey contours), compared
to the parameter $b_{1}$, estimated from the 2PCF (red vertical
line). Specifically, the light and grey contour plots are the $68$ and
$95$ per cent likelihood probability contours in the
$b_{1}-b_{2}/b_{1}$ planes, obtained with Eqs.~\ref{eq:bias3PCF} and
\ref{eq:chi} and combining together the measurements at the three
scales considered here.

Figure~\ref{fig:bias_nocov2} extends the previous analysis, showing
the halo bias in the redshift range $0\leq z\leq2$ and comparing the
$\Lambda$CDM case with the EXP003 and SUGRA003 simulations, as
indicated by the labels. Among the cDE models of the {\small CoDECS}
simulations, EXP003 and SUGRA003 are the ones that differ the most
from the $\Lambda$CDM case. Coloured and black lines show the results
obtained with separate and joint scales, respectively. These
estimates are compared with both the biases obtained from the 2PCF
(through Eq. \ref{eq:bias2PCF}), in the range $10<r<50$ \Mpch, and
with the theoretical $\Lambda$CDM effective bias values predicted by
\citet{sheth_mo_tormen2001} and \citet{tinker2010}, normalised to the
$\sigma_{8}$ values of each respective model, and weighted by the halo
mass function, as in MBM12 (see their Fig.2 and related
discussion). The shaded areas show a representative $10\%$ scatter,
that reflects the uncertainty in the theoretical predictions, as found
in the literature \citep[e.g. see the differences
between][]{sheth_tormen1999,sheth_mo_tormen2001,tinker2010}. In
the lower panels of Fig.~\ref{fig:bias_nocov2}, we show the 
fractional difference in percent of the bias as estimated from the
3PCF with respect to the bias estimated from the 2PCF, i.e. 
$\Delta b/b_{2PCF}[\%]=(b_{3PCF}-b_{2PCF})/b_{2PCF}\cdot100$.

While the linear bias obtained from the 2PCF results in excellent
agreement with theoretical predictions, at all redshifts and for all
the cDE models considered, as already noted by MBM12, the goodness of
the bias derived from the 3PCF depends on the scales used. In
particular, the agreement is better when the bias is estimated from
the reduced 3PCF measured at the largest scales ($r_{12}\geq 10$ \Mpch), as
expected since the linear bias model given by Eq.~\ref{eq:bias3PCF} is
a good approximation only in the linear regime ($r\gtrsim10$
\Mpch). On the other hand, the larger discrepancies at higher redshift
are due to the sparseness of DM samples used to estimate $Q_{\rm DM}$
in Eq.~\ref{eq:bias3PCF}. A more detailed analysis using theoretical
predictions for $Q_{\rm DM}$ is deferred to a future paper.

Figure~\ref{fig:bias_nocov2} shows that the bias estimated from the
combination of the reduced 3PCF measured at all three scales
considered better reproduces the real bias of our simulations, even
though in this analysis we do not take into account the covariance
between different scales (see discussion in Appendix
\ref{sec:cov}). We note, however, that at high redshift this is just a
spurious effect, caused by the compensation of the underestimation of
the bias at small scales and the overestimation at large scales.

As already noted in Fig.~\ref{fig:biasLCDM_z1}, the bias estimated
from the 3PCF appears in reasonable agreement with the true one at
$z=1$, especially when assessed from the combination of
scales. Moreover, Fig.~\ref{fig:bias_nocov2} shows that this is true
also for the models EXP003 and SUGRA003. On the other hand, the
agreement is worst at both lower and higher redshifts. In particular,
similarly as previous findings by \citet{marin2008}, our analysis
shows that the bias estimated from Eq.~\ref{eq:bias3PCF} slightly
overestimates the one given by Eq.~\ref{eq:bias2PCF} by $\sim
15-25\%$, on average, as can be seen in the lower panels of
Fig.~\ref{fig:bias_nocov2}. However, we point out that the
significance of such a discrepancy, that in many cases is below the
$1\,\sigma$ estimated uncertainty, may be further reduced when
considering also the error on the bias estimate from the 2PCF,
typically of the order of 5-10 $\%$ (e.g. see the red shaded area of
Fig. \ref{fig:biasLCDM_z1_cont}). These discrepancies become
smaller, of the order of $\sim 10-20\%$, and similar to the ones
quoted by \citet{marin2008}, when considering the analysis at combined
scales. Our results are also in agreement with the work by
\citet{hoffmann2014}. They analysed the MICE-GC $\Lambda$CDM
simulation with a method similar to the one adopted in this paper,
finding that the linear bias estimated from the 3PCF $Q$
overestimates the one from the 2PCF by $\sim$30-40\%, at all mass
and redshift ranges considered. Our work extends these findings to
cDE cosmologies.

Overall, these systematics indicate that assuming a local
deterministic bias model can significantly affect the results, and
that further developments from the theoretical side are required.


\begin{figure}
\includegraphics[angle=0, width=0.5\textwidth]{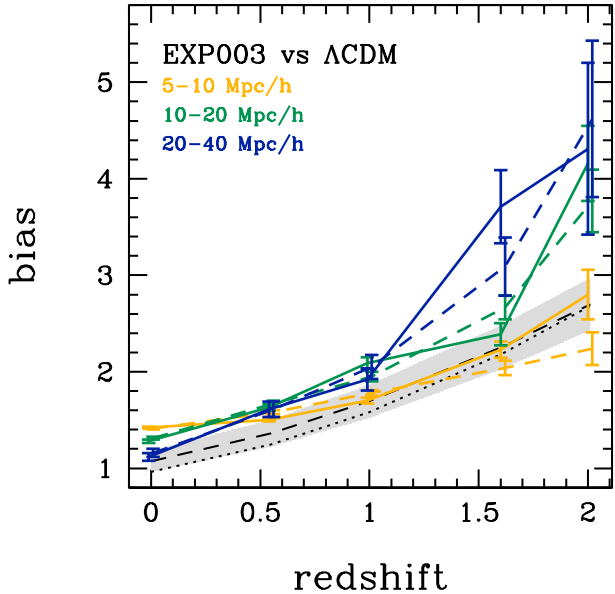}
\caption{Bias of CDM haloes estimated from the 3PCF, using different
cosmological models to compute $Q_{\rm h}$ and $Q_{\rm DM}$. In particular, 
$Q_{\rm h}$ is measured in the EXP003 simulation,
while $Q_{\rm DM}$ in the $\Lambda$CDM one. As a comparison, we also reported the bias function
obtained with the correct cosmology (dashed coloured lines) and
theoretical predictions (dotted and dashed lines). The symbols of
the latter quantities are the same as in
Fig.~\ref{fig:bias_nocov2}. The points have
been slightly shifted along the horizontal axis to make the figure clearer.}
\label{fig:bias_diagcov_mix2}
\end{figure}

\subsection{Geometric distortions}
\label{sec:real-analysis}

In the previous section we demonstrated that the 3PCF can be used to
estimate the halo bias in a $\sigma_{8}$-independent way. All the
measurements discussed so far assumed the correct underlying cosmology
when estimating both the CDM and the halo correlation
functions, i.e. we always employed cosmic distances that were numerically 
computed according to each cosmological model under investigation. 
However, when analysing real datasets, the true cosmology
of the Universe is unknown.

If a wrong cosmology is assumed when converting redshifts into
comoving distances, the measured 2PCF and 3PCF will be distorted. As
discussed in MBM12, this geometric effect is small for the cDE models
analysed in this work, since they are nearly indistinguishable in
terms of the Hubble function. Besides this effect on measured
quantities, a wrong assumption on the underlying cosmology impacts
also the function $Q_{\rm DM}(\theta)$ used in
Eq.~\ref{eq:bias3PCF}. The latter is generally estimated using
standard $\Lambda$CDM simulations, with fixed cosmological parameters
\citep[see e.g.][]{gaztanaga2005b, ross2006, mcbride2011, mcbride2011b, marin2011,
 marin2013}. At the present state, no attempt has been done to
investigate the impact of assuming a wrong cosmology when estimating
$Q_{\rm DM}(\theta)$, especially in non-standard cosmological
scenarios.

In the following, we test this effect in the case of cDE models using
our {\small CoDECS} datasets. Specifically, we investigate the effect
of measuring the bias from Eq.~\ref{eq:bias3PCF} assuming different
models for $Q_{\rm h}$ and $Q_{\rm DM}$. To maximise the effect, we
consider the $\Lambda$CDM and EXP003 cosmologies, since the latter is
the model that differs the most from the $\Lambda$CDM case.
Fig.~\ref{fig:bias_diagcov_mix2} displays the results of this test. In
this figure, we show the bias of CDM haloes estimated using the
function $Q_{\rm h}(\theta)$ measured in the EXP003 simulations,
while $Q_{\rm DM}(\theta)$ is measured in the $\Lambda$CDM ones. This case
represents the hypothetical situation in which an observer living in a
(quite extreme) cDE universe attempts to estimate the bias assuming a $\Lambda$CDM
model to compute $Q_{\rm DM}(\theta)$. As it can be seen, the bias functions
obtained with (dashed lines) and without (solid lines) assuming the
correct cosmology for the DM 3PCF are consistent, considering the
quoted errorbars, for all the scales and redshift analysed, with no
particular systematic trend.
This is a quite remarkable result: the determination of the bias
through the 3PCF seems to be mostly independent of the cosmological 
model, even for rather complex and extreme extensions beyond the 
standard $\Lambda $CDM scenario.


\section{Conclusions} 
\label{sec:concl}

In this work we investigated the power of the 3PCF to constrain cDE
models, and to estimate the halo bias in these cosmological scenarios. 
We analysed halo catalogues extracted from
the {\small CoDECS} simulations, which are
large, collisionless N-body simulations of a variety of cosmological
models in which the DE scalar field can interact with CDM particles by
exchanging energy-momentum. 
We estimated the 3PCF $\zeta(\theta)$ and
the reduced 3PCF $Q(\theta)$ for triplets of objects in which
$r_{12}/r_{13}=2$, where $r_{12}$ and $r_{13}$ are the comoving
separations between two of the objects, and $\theta$ is the angle
between them. We considered different representative scales
($r_{12}=2, 3, 5, 10, 20$ \Mpch) and redshifts ($z=0, 0.55, 1, 1.61,
2.01$).

The main results of this analysis, independently of the cosmological
model considered, are as follows:
\begin{itemize}
\item at both fixed scale and redshift, the reduced 3PCF is higher
for elongated triangles than for perpendicular ones, as
expected from non-linear gravitational instability theory;
\item at fixed scale, the 3PCF increases going to higher redshift,
while the reduced 3PCF becomes flatter, which is a signature of the evolution
of the bias function and of the formation of filaments with cosmic
time;
\item at fixed redshift, we see a transition in the reduced 3PCF from
a {\it U-shape}, at small scales, to a {\it V-shape}, at large
scales. This is a consequence of the fact that, at small scales,
structures preferentially reside in approximately round structures (hence
presenting a flatter $Q$), while at larger scales also the
contribution of structures in filaments starts to become significant.
\end{itemize}
Overall, these results confirm what already found in the literature at
$z\sim0$ \citep[e.q.][]{gaztanaga2005, marin2008}, and extend it to a
wider redshift range and different cosmologies.

Regarding the effects of DE coupling on the halo 3PCF, our main results
are as follows:
\begin{itemize}
\item the 3PCF in cDE models is significantly different than in the
$\Lambda$CDM scenario, with deviations that increase going to larger
scales;
\item at fixed scale, cDE models predict a higher (smaller) 3PCF for
elongated (perpendicular) configurations with respect to the
$\Lambda$CDM cosmology;
\item the differences increase with redshift, being almost negligible
at $z=0$.
\end{itemize}

These 3PCF measurements have been used to constrain the linear bias
function of CDM haloes in the redshift range $0\leq z\leq2$. We
considered three configuration scales ($s=5, 10, 20$ \Mpch), both
separately and with a joint analysis. We find a reasonable agreement,
considering the estimated uncertainties, between the bias estimated
from the 3PCF and from the 2PCF. However, we also find that the
3PCF tends to systematically overestimate the bias by $\sim$15-25\%
on average, mainly due to the linear approximations in the
theoretical modelisation. These results are in good agreement with
the works of \citet{marin2008} and \citet{hoffmann2014}, for the
$\Lambda$CDM case. Finally, we quantified the impact of assuming a
wrong cosmology in estimating the halo bias, using different
cosmologies for the reduced 3PCF of DM and haloes. Even in the most
extreme case analysed in this work, we find that this effect is
negligible considering the uncertainties.

This work demonstrates that the 3PCF can be efficiently exploited, as
a complementary probe to lower-order statistics, to discriminate
between alternative cosmological scenarios. In particular, we proved
that the degeneracy between the halo bias and $\sigma_{8}$ can be
broken using the reduced 3PCF $Q(\theta)$. Moreover, combining
measurements of redshift-space distortions in the 2PCF with
$\sigma_{8}$ constraints from the reduced 3PCF, it will be possible to
estimate directly the growth rate of structures $f(z)$.

The differences found between the various models are indeed small, and
the errors on the bias provided by present surveys are just of the
order of these differences, not yet allowing to disentangle between
standard and alternative cosmologies \citep[e.g.,
see][]{marin2013}. However, ongoing and future surveys, such as BOSS
\citep{boss}, WiggleZ \citep{blake2011a}, VIPERS \citep{guzzo2013},
and Euclid \citep{euclid}, will be able to increase current statistics
by more than one order of magnitude, reducing dramatically the
expected statistical errors, though a more detailed
assessment of the theoretical framework is required to understand
the systematics highlighted in this work. In this context, the
analysis of the 3PCF will be of extreme importance also at the BAO
scale, as recently shown by \citet{fosalba2013} and
\citet{crocce2013}.

\section*{acknowledgments}
We would like to thank the anonymous referee for the useful
comments and suggestions, which helped to improve the quality of the
paper. We acknowledge financial contributions by grants ASI/INAF
I/023/12/0, PRIN MIUR 2010-2011 ``The dark Universe and the cosmic
evolution of baryons: from current surveys to Euclid'' and PRIN INAF
2012 ``The Universe in the box: multiscale simulations of cosmic
structure''. MB is supported by the Marie Curie Intra European
Fellowship ``SIDUN" within the 7th Framework Programme of the European
Commission.

\bibliographystyle{mn2e} \bibliography{bib}

\appendix 
\section{Covariance matrices}
\label{sec:cov}

\begin{figure*}
\includegraphics[angle=0, width=0.9\textwidth]{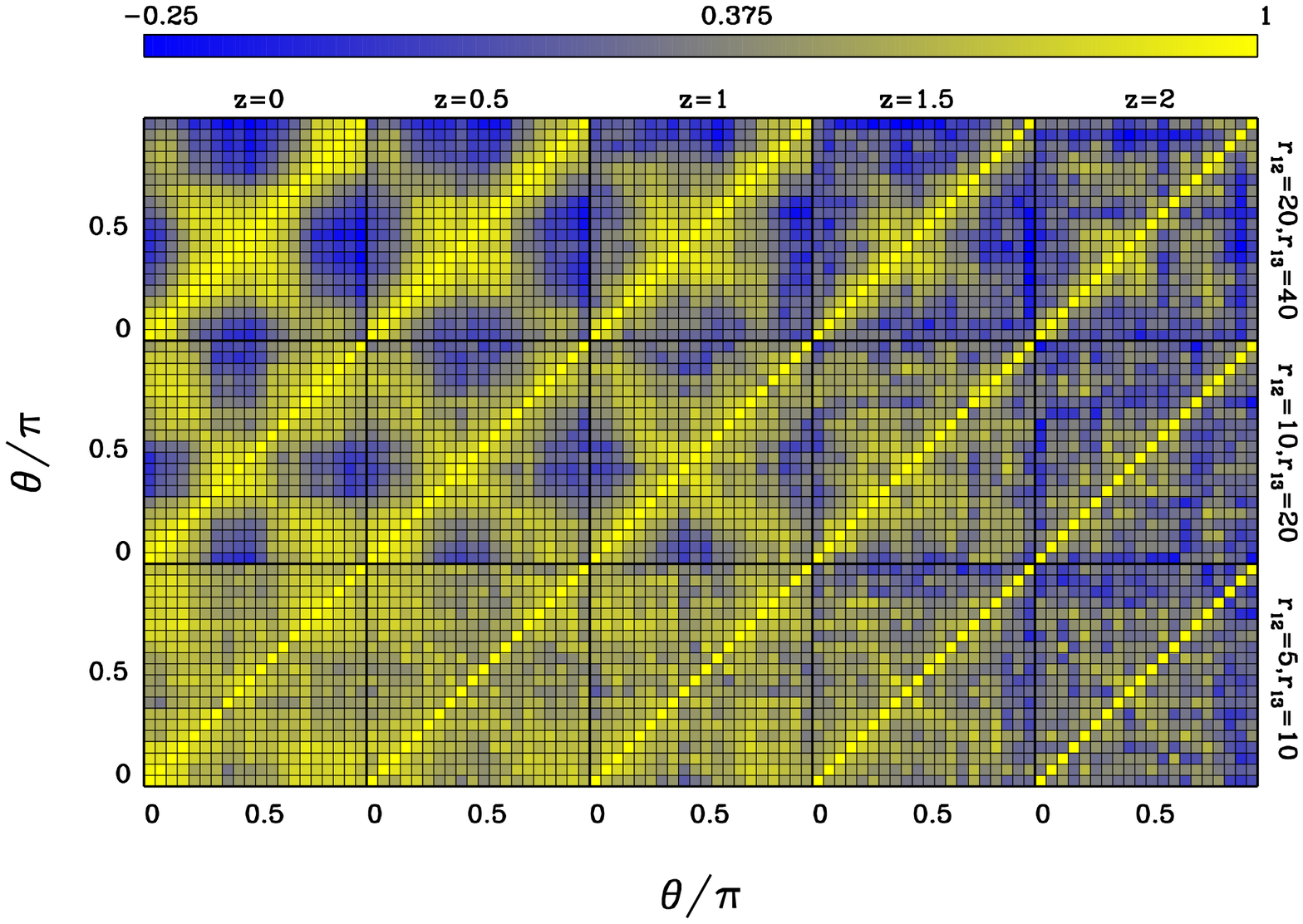}
\caption{Normalised covariance matrices estimated with the large $\Lambda$CDM simulation 
of the new {\small XL-CoDECS} series ($2\times 2048^{3}$ particles over a $2$ 
\Gpch periodic cosmological box), at all scales (from lower to upper panels) and 
redshifts (from left to right) considered in this work. The redshift range is labeled 
above the upper panels, while at the right can be found the scale considerend.}
\label{fig:matrices_all}
\end{figure*}

\begin{figure*}
\begin{center}
\includegraphics[angle=0, width=0.45\textwidth]{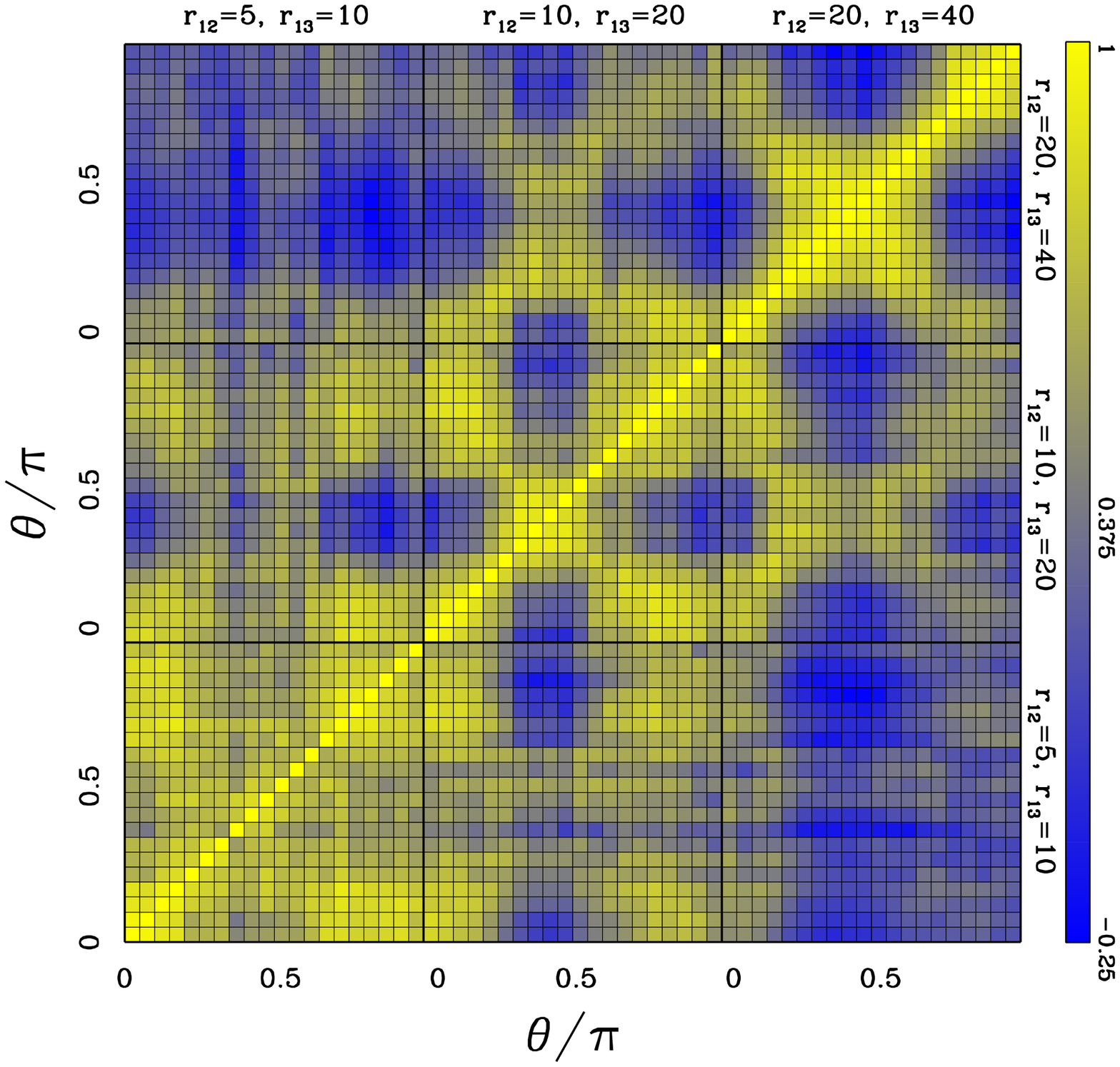}
\includegraphics[angle=0, width=0.45\textwidth]{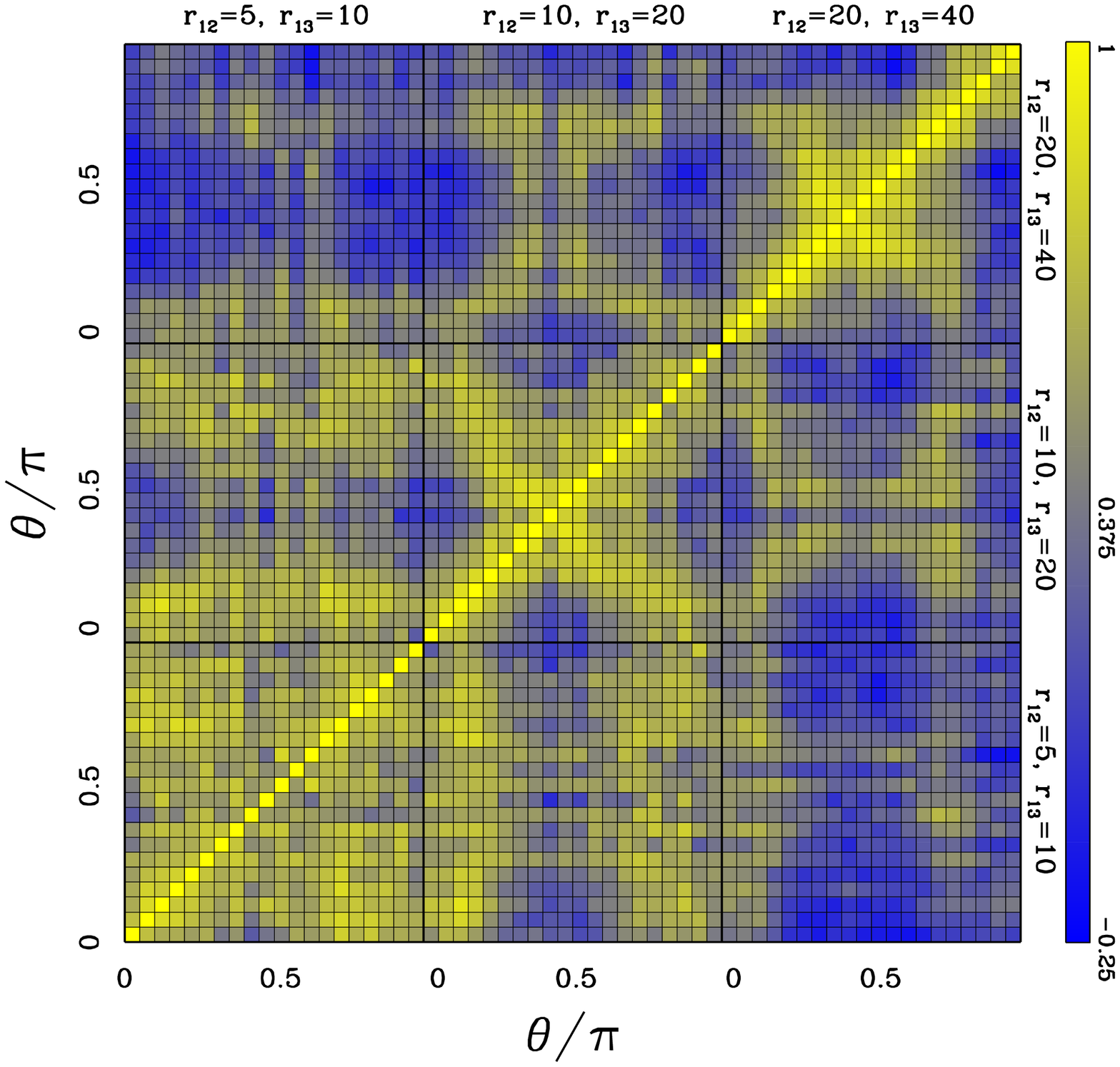}
\caption{Normalised covariance matrices between mixed scales, at redshifts $z=0$
(left panel), $z=1$ (right panel). The diagonal
squares presents just the normalised covariance matrix of every single scale, as
shown in Fig.~\ref{fig:matrices_all}, while the off-diagonal squares present the
normalised covariance matrix between different scales; the corresponding 
scale can be found labeled above and at the right of each panel.}
\label{fig:matrices_mix}
\end{center}
\end{figure*}

\begin{figure}
\begin{center}
\includegraphics[angle=0, width=0.45\textwidth]{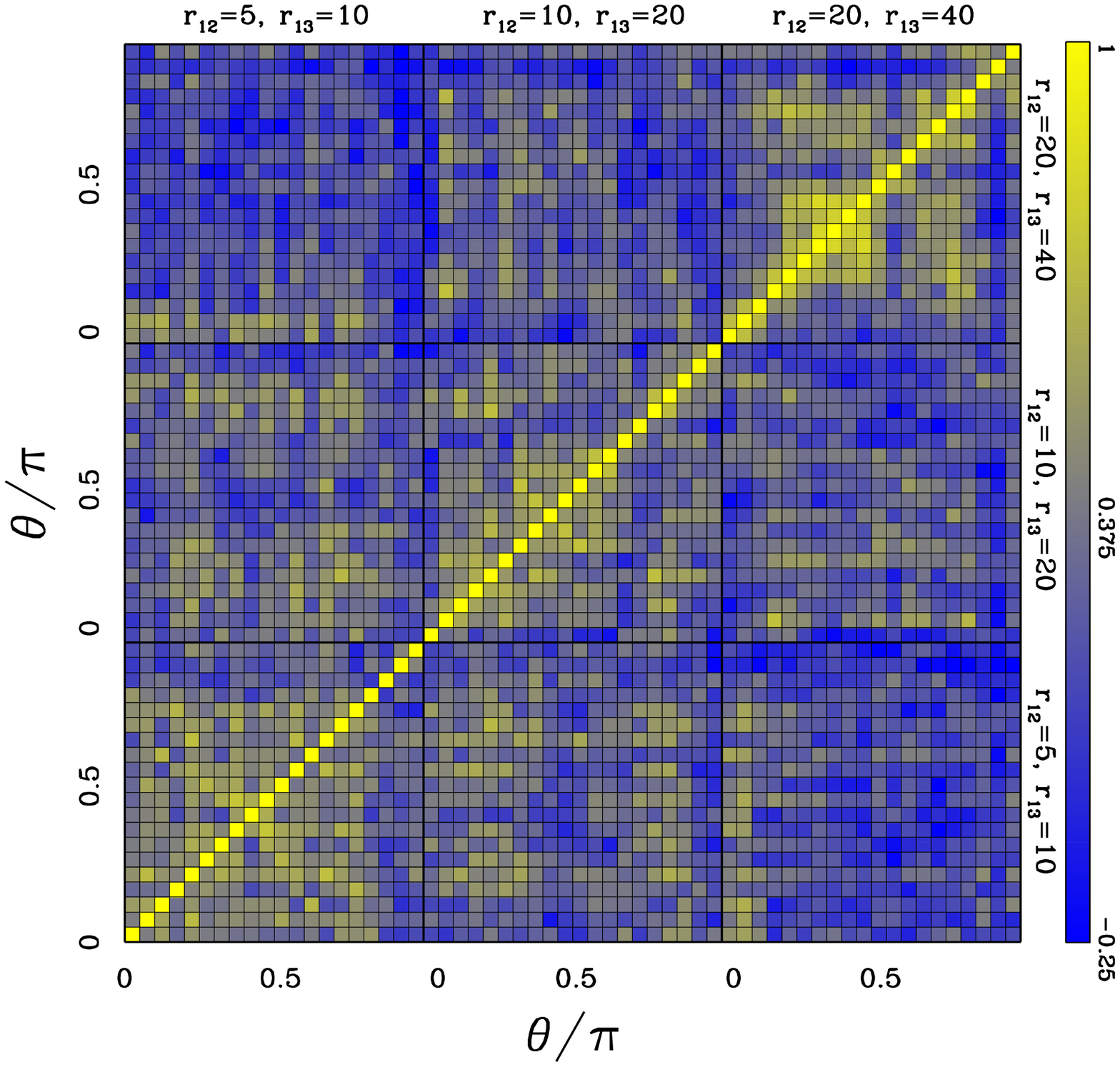}
\caption{Normalised covariance matrices between mixed scales, at
redshift $z=2$. The labels are the same as
Fig.~\ref{fig:matrices_mix}}
\label{fig:matrices_mix1}
\end{center}
\end{figure}

The covariance matrices used in this paper to estimate the errors on
$\zeta(\theta)$ and $Q(\theta)$ are derived using the large {\small
XL-CoDECS} $\Lambda$CDM simulation. As explained in \S
\ref{sec:growth}, we divided the {\small CoDECS-XL} snapshots into
$N=27$ sub-cubes of side $\sim 700$ \Mpch. The normalised covariance
matrix is defined as follows:
\begin{equation}
\displaystyle C_{ij} =
\frac{1}{N}\sum_{k=1}^{N}\left(\frac{Q_{i}^{k}-\bar{Q}_{i}}{\sigma_{Q_{i}}}\right)
\left(\frac{Q_{j}^{k}-\bar{Q}_{j}}{\sigma_{Q_{j}}}\right) \, ,
\end{equation}
where $\bar{Q}_{j}$ is the mean value of the reduced 3PCF in the $N$
sub-cubes. Fig. \ref{fig:matrices_all} shows the normalised
covariance matrices at all the redshifts and scales considered in this
analysis, i.e. $0\leq z\leq2$, $s=2$ \Mpch and $u= 5, 10, 20$
\Mpch. We find a significant covariance between different bins,
especially at small angles and scales. At high redshifts and large
scales, the covariance matrices become gradually more and more
diagonal. Qualitatively, these results are in good agreement with the
covariance matrices found for similar configurations by \citet{gaztanaga2005}
and \citet{hoffmann2014}.
To perform the joint analysis discussed in \S
\ref{sec:real-analysis}, we also estimated the covariance matrix
between different scales. The result is shown in
Figs.~\ref{fig:matrices_mix} and \ref{fig:matrices_mix1}, for three characteristic redshifts. At
high redshift, the covariance between different scales is quite small,
and the matrix is almost diagonal. The correlations between scales
increases with decreasing redshift.

To derive the bias parameters of Eq.~\ref{eq:bias3PCF}, we performed a
full-covariance $\chi^{2}$ analysis, with:
\begin{equation}
\displaystyle \chi^{2} = \sum_{i=1}^{N}\sum_{j=1}^{N}
\left(\frac{Q_{i}(\theta)-Q^{\rm
  model}_{i}(\theta)}{\sigma_{Q_{i}(\theta)}}\right) C_{ij}^{-1}
\left(\frac{Q_{j}(\theta)-Q^{\rm
  model}_{j}(\theta)}{\sigma_{Q_{j}(\theta)}}\right) \, .
\end{equation}
However, in most cases the limited number of available sub-mocks
introduces numerical instabilities in the inversion of $C_{ij}$,
biasing the final results. For this reason we decided to use only the
diagonal elements of the covariance matrix in our computations (see
Eq.~\ref{eq:chi}). For a small subset of cases where this issue
was less severe, we verified that the difference in the bias
parameter estimation is of the order of 0.1, and the difference in
the estimated error of the order of 0.015. A similar analysis of the
impact of the covariance matrix on the bias estimate has been
carried out also by \citet{hoffmann2014}, reaching a similar
conclusion that the results are not strongly dependent on the
uncertainty in the covariance matrix. 

Moreover, by looking at Fig.~\ref{fig:matrices_all}, we see that the
impact of this assumption is most severe at low redshifts and at small
scales. On the contrary, at large scales and redshifts the
off-diagonal terms are small, and can be safely neglected.

\end{document}